\documentstyle[psfig]{mn05}

\def\simg{\mathrel{%
      \rlap{\raise 0.511ex \hbox{$>$}}{\lower 0.511ex \hbox{$\sim$}}}}
\def\siml{\mathrel{%
      \rlap{\raise 0.511ex \hbox{$<$}}{\lower 0.511ex \hbox{$\sim$}}}}
\def\eq{eq$.$} \def\eqs{eqs$.$} \def\etal{et al$.$ } \def\eg{e$.$g$.$ } \def\ie{i$.$e$.$ }
\def\Da{\Delta \alpha} \def\a1{\alpha_1} \def\hleft{\hspace*{-2mm}} \def\hLeft{\hspace*{-3mm}}

\def\reference{\bibitem}

\begin{document}

\title[ Jets, Structured Outflows, and Energy Injection in GRB Afterglows]
      { Models for Achromatic Light-Curve Breaks in GRB Afterglows: Jets, Structured Outflows, and Energy Injection }

\author[A. Panaitescu]{A. Panaitescu \\
        Department of Astronomy, University of Texas, Austin, TX 78712}

\maketitle

\begin{abstract}

 The steepening (break) of the power-law fall-off observed in the optical emission of some GRB afterglows 
 at epoch $\sim 1$ day is often attributed to a collimated outflow (jet), undergoing lateral spreading.
 Wider opening GRB ejecta with a non-uniform energy angular distribution (structured outflows) or the 
 cessation of energy injection in the afterglow can also yield light-curve breaks.  \\
 We determine the optical and $X$-ray light-curve decay indices and spectral energy distribution slopes 
 for 10 GRB afterglows with optical light-curve breaks (980519, 990123, 990510, 991216, 000301, 000926, 
 010222, 011211, 020813, 030226), and use these properties to test the above models for light-curve steepening. 
 It is found that the optical breaks of six of these afterglows can be accommodated by either energy injection 
 or by structured outflows. In the refreshed shock model, a wind-like stratification of the circumburst medium 
 (as expected for massive stars as GRB progenitors) is slightly favoured. A spreading jet interacting with a 
 homogeneous circumburst medium is required by the afterglows 990510, 000301, 011211, and 030226. 
 The optical pre- and post-break decays of these four afterglows are incompatible with a wind-like medium. \\
 The current sample of 10 afterglows with breaks suggests that the distribution of the break magnitude $\Da$ 
 (defined as the increase of the afterglow decay exponent) is bimodal, with a gap at $\Da \simeq 1$. If true, 
 this bimodality favours the structured outflow model, while the gap location indicates a homogeneous circumburst 
 environment.  

\end{abstract}
  

\section{Introduction}

 The optical light-curves of most afterglows monitored over more than one decade in time exhibit 
a steepening at about 1 day after the GRB. Such a break has been predicted by the sideways spreading
jet model of Rhoads (1999) and was observed shortly afterward for the first time in the afterglow 
990123 (Kulkarni \etal 1999). Since then, many other afterglows displayed an optical light-curve 
break, to which the jet model has been extensively applied. 

 Rees \& M\'esz\'aros (1998) have proposed that the blast wave may be "refreshed" when there
is a substantial energy input in the forward shock. Such an energy injection could occur either 
if there is a wide distribution of Lorentz factors in the GRB explosion, the slower ejecta catching 
up with those which were initially faster during their deceleration by the circumburst medium, 
or if the central energy is long lived, releasing a relativistic outflow for hours or days after 
the explosion. The increase of the ejecta energy mitigates the deceleration of the forward shock, 
which results in a slower decay of the afterglow flux, as was proposed by Fox \etal (2003) to explain
the nearly flat optical light-curve of the afterglow 021004 at 0.003--0.1 days after the burst. 
In this model, a steepening of the afterglow decay would occur when the rate of energy injection 
decreases.

 Another possibility, proposed by M\'esz\'aros, Rees \& Wijers (1998), is that the angular
distribution of the ejecta kinetic energy is not uniform. Then, as the fireball decelerates, 
the relativistic beaming of the radiation allows the observer to receive emission from an ever 
wider region of the fireball, whose energy varies with observer time differently than in the 
case of an isotropic energy distribution. In this model, a light-curve break would result
if the second order angular derivative of the energy per solid angle is negative, as in the
scenario proposed by Rossi, Lazzati \& Rees (2002) and Zhang \& M\'esz\'aros (2002).

 The purpose of this paper is to test the above possible models for afterglow beaks: refreshed 
shocks, structured outflows, and spreading jets. To this end, we compare the light-curve breaks 
expected in each model with those measured in the optical emission of ten GRB afterglows, and add 
consistency checks with the $X$-ray light-curves and the optical-to-$X$-ray spectral energy
distribution (SED) slope.

\section{Optical Continuum Slope and Light-Curve Indices}
\label{models}

 The simplest and most widely used test of the blast wave model for GRB afterglows consists of 
comparing the light-curve decay index with the SED slope. It is a robust prediction of the 
standard blast wave model (M\'esz\'aros \& Rees 1997), confirmed by observations (\eg Wijers, Rees
\& M\'esz\'aros 1997), that the afterglow light-curve at a frequency $\nu$ above the peak of the 
afterglow spectrum decays as a power-law
\begin{equation}
  F_\nu \propto t^{-\alpha} \;,
\label{fnu}
\end{equation}
where $\alpha$ depends only on the location of the cooling frequency ($\nu_c$) relative to $\nu$,
the slope $p$ of the power-law distribution $dN/d\epsilon$ with the energy $\epsilon$ of the 
electrons accelerated by the blast-wave
\begin{equation}
 dN/d\epsilon \propto \epsilon^{-p} \;,
\label{dNde}
\end{equation}
and the radial stratification of the circumburst medium:
\begin{equation}
 n(r) \propto r^{-s} \;,
\label{n}
\end{equation}
with $s=0$ for a homogeneous environment and $s=2$ for the free wind of a massive GRB progenitor.
For a fireball (or a jet at times when it is sufficiently relativistic that relativistic beaming 
prevents its finite angular opening to affect the observed afterglow emission), the light-curve 
index of equation (\ref{fnu}) is given by 
\begin{equation}
 \alpha = \frac{1}{4} \cdot  \left\{ \begin{array}{ll}  3p-3 \;, & \nu < \nu_c \;\; \& \;\; s=0 \\ 
           3p-2 \;, & \nu_c < \nu \;\; \& \;\; s=0,2 \\  3p-1 \;, & \nu < \nu_c  \;\; \& \;\; s=2  
           \end{array} \right.  \;
\label{alpha}
\end{equation}
(Sari, Piran \& Piran 1998). The second line in the $rhs$ of equation (\ref{alpha}) assumes that 
electron cooling is dominated by synchrotron losses. If radiative losses occur mainly through 
inverse Compton scatterings, then  
\begin{equation}
 \alpha (\nu > \nu_c) = \frac{3}{4} p - \left\{ \hleft \begin{array}{llll} 
           (4-p)^{-1} &,& s=0 \;\; \& \;\; 2 < p < 3 \\  1   &,& s=0 \;\; \& \;\; 3 < p \\ 
            p/(8-2p)  &,& s=2 \;\; \& \;\; 2 < p < 3 \\  1.5 &,& s=2 \;\; \& \;\; 3 < p 
   \end{array} \right.  
\label{nuc}
\end{equation}
(Panaitescu \& Kumar 2001).
We ignore the possibility of inverse Compton dominated electron cooling, keeping in mind that, 
because of this  deficiency of our treatment, we may miss some viable models with $\nu_c < \nu$.

For the electron distribution given in equation (\ref{dNde}), the intrinsic afterglow SED 
at optical and $X$-ray frequencies is a power-law:
\begin{equation}
 F_\nu \propto \nu^{-\beta} \;, \quad \beta = \frac{1}{2} \cdot \left\{ \begin{array}{lll} 
               p-1 & , & \nu < \nu_c \\ p & , &  \nu_c < \nu  \end{array} \right. \;.
\label{beta}
\end{equation}
Eliminating the electron index $p$ between equations (\ref{alpha}) and (\ref{beta}), leads to
\begin{equation}
 3 \beta - 2 \alpha = 0, 1, -1 \;,
\label{ab}
\end{equation}
where the order in the $rhs$ is the same as in equation (\ref{alpha}). 
Equation (\ref{ab}) is often used to test the blast wave model and to identify the type of external 
medium (if $\nu < \nu_c$). However, this exercise may not always provide a conclusive result 
because a small amount of dust extinction in the host galaxy can alter significantly the slope
of the optical SED. For an extinction of $A_{(z+1)\nu}$ magnitudes at an observer frequency $\nu$, 
the local slope of the reddened SED is
\begin{displaymath}
 \beta (\nu) = - \frac{d}{d \ln \nu} \ln [F_\nu \cdot 10^{-0.4 A_{(z+1)\nu}}] =
\end{displaymath}
\begin{equation}
  \quad \quad  \quad  = \beta_o + 0.9\, A_{(z+1)\nu} \frac{d \ln A_\nu}{d \ln \nu} \;,
\end{equation}
where $\beta_o$ is the intrinsic (unreddened) SED slope.
For a simple $A_\nu \propto \nu$ reddening curve,
\begin{equation}
 \beta (\nu) = \beta_o + 0.9\, (z+1)\, A_\nu \;,
\label{betaobs}
\end{equation}
where $A_\nu$ is the extinction in the host frame at the observing frequency $\nu$. Hence,
a host extinction of $A_V \simeq 0.1$ mag is enough to confuse the test based on equation 
(\ref{ab}) for an afterglow at redshift $z \geq 1$.

 There are a few mechanisms (described below) which can alter the light-curve decay index 
$\alpha$ of equation (\ref{alpha}) and, hence, modify the test equation (\ref{ab}).

\subsection{Energy Injection}

 As proposed by Rees \& M\'esz\'aros (1998), the blast wave may be "refreshed" by some delayed
(or lagged) ejecta. For convenience, we quantify the energy injection by expressing the fireball 
energy $E$ as a function of the observer time $t$. That afterglow light-curves are power-laws 
in observer time indicates that $E(t)$ is not far from a power-law either. For this reason, 
we consider that 
\begin{equation}
  E(t) \propto t^e \;.
\label{e}
\end{equation}
Assuming negligible radiative losses after about 0.1 day, when afterglow observations are usually
made, the fireball Lorentz factor is given by $\Gamma^2 M(r) \propto E$, where $M(r) \propto r^{3-s}$ 
is the mass of the swept-up circumburst medium. Together with $r \propto \Gamma^2 t$, equation 
(\ref{e}) is equivalent with the following distribution of the ejecta energy with Lorentz factor:
\begin{equation}
  \frac{dE}{d\Gamma} \propto \Gamma^{-\gamma} \;, \quad 
     \gamma = \left\{ \begin{array}{ll}  (7e+3)/(3-e)\;, & s=0  \\  
                      (3e+1)/(1-e)\;, & s=2 \end{array}  \right. \;,
\label{gamma}
\end{equation} 
for $e < 3$ ($e < 1$) for a homogeneous (wind-like) medium.
The effect of energy injection on the afterglow flux $F_\nu$ decay can be easily assessed from 
the expression of $F_\nu(t)$ as function of the fireball energy $E$ for the no-injection case. 
 From equations (B7), (B8), and (C5) of Panaitescu \& Kumar (2000), the effect of energy 
injection is a reduction of the light-curve decay index given in equation (\ref{alpha}) by
\begin{equation}
 \Da_{ei} =  \frac{e}{4} \cdot \left\{ \begin{array}{ll} 
                 p+3 \;, & \nu_o < \nu_c \;\&\; s=0 \\ p+2 \;,& \nu_c < \nu_o \;\&\; s=0,2  \\
                 p+1 \;, & \nu_o < \nu_c \;\&\; s=2    \end{array} \right.  \;.
\label{EI}
\end{equation} 
When the injection of energy subsides, the power-law afterglow decay steepens from 
$t^{-(\alpha - \Da_{ei})}$ to $t^{-\alpha}$, assuming that the ejecta are spherical 
or sufficiently relativistic that the effects associated with collimation are not yet manifested.

\subsection{Structured Outflows}

 Another possibility is that the ejecta energy per solid angle, $dE/d\Omega$, is not constant, 
but varies with the angle $\theta$ measured from the outflow symmetry axis (M\'esz\'aros, Rees 
\& Wijers 1998). The acceleration of the GRB outflow and penetration through what is still left 
of the progenitor stellar envelope should lead to a $dE/d\Omega$ decreasing with increasing 
polar angle $\theta$, as illustrated in fig. 11 of MacFadyen, Woosley \& Heger (2001), given 
that the mass per solid angle of the helium core increases with $\theta$ (hence a higher binding 
energy and a larger dissipation of the jet energy, leading to a faster jet spread, at larger angles). 
That afterglow light-curves are power-laws in time suggests that $dE/d\Omega$ can be sufficiently 
well approximated as a power-law in $\theta$: 
\begin{equation}
  dE/d\Omega \propto \left\{ \begin{array}{lll} const &,& \theta < \theta_c \\ 
                (\theta/\theta_c)^{-q} &,& \theta > \theta_c \end{array} \right.\;,
\label{q}
\end{equation}  
with a uniform core of opening $\theta_c$ to avoid the on-axis divergence. To make use
of analytical results for the light-curve decay index (Panaitescu \& Kumar 2003), we assume 
in this work that the direction toward the observer lies within the core opening. In this
case, when the blast-wave has decelerated enough that the emission from the outflow envelope 
is beamed relativistically toward the observer (\ie when the fireball Lorentz factor $\Gamma$ 
drops below $\theta_c^{-1}$), the afterglow emission decay steepens from that given in equation 
(\ref{alpha}) to $F_\nu \propto t^{-(\alpha + \Da_{so})}$, where
\begin{equation}
 \Da_{so} = \left\{ \begin{array}{lll} 
       \frac{\displaystyle 3q(8-\tilde{q})}{\displaystyle 4\tilde{q}(8-q)}  & , & s=0  \\
       \frac{\displaystyle  q(4-\tilde{q})}{\displaystyle 2\tilde{q}(4-q)}  & , & s=2 
     \end{array} \right. \;,
\label{SO}
\end{equation}
because the envelope has an energy density smaller than that of the core. Equation (\ref{SO}) holds
provided that $q < \tilde{q}$, with
\begin{equation}
  \tilde{q} = \left\{ \begin{array}{ll}
    \frac{\displaystyle 8}{\displaystyle p+4}\;, & (s=0, \nu < \nu_c) \;\; {\rm or} \;\; (s=2, \nu_c < \nu) \\ 
    \frac{\displaystyle 8}{\displaystyle p+3}\;, & (s=0, \nu_c < \nu) \;\; {\rm or} \;\; (s=2, \nu < \nu_c)  
        \end{array} \right. \;.
\label{qt}
\end{equation}
If the observer lies outside the core, the pre-break decay index cannot be obtained analytically.
Obviously, as the more energetic core becomes visible to the observer, the afterglow decay should
be shallower than that given in equation (\ref{alpha}) and the break magnitude larger than that of 
equation (\ref{SO})  (see fig. 2 and 3 in Panaitescu \& Kumar 2003).

 For $q > \tilde{q}$ the ejecta energy density $dE/d\Omega$ falls-off away from the axis sufficiently 
fast that the afterglow emission is dominated by the core. In this case, when the core boundary becomes 
visible the afterglow decay index increases by
\begin{equation}
  \Da_{j/o} = \left\{ \begin{array}{ll}  3/4\;, & s=0 \\ 1/2\;, & s=2 \end{array}\right. \;,
\label{JO}
\end{equation}
independent of frequency (Panaitescu, M\'esz\'aros \& Rees 1998).
The subscript "j/o" stands for "jet in an outflow" and hereafter will identify the model of a jet 
whose lateral spreading is prevented by the envelope but which gives essentially all the afterglow 
emission.

\subsection{Jets}

 An even larger break $\Da$ than given in equation (\ref{JO}) is obtained if the envelope has too 
little ejecta to impede the lateral spreading of the core, as the sideways expansion of the jet 
leads to a faster decrease of its Lorentz factor, starting at about the same time when the jet 
edge becomes visible. This model represents the extreme case of a structured outflow with $dE/d\Omega$ 
a "top-hat" function (\ie $q \rightarrow \infty$ in \eq [\ref{q}]). When the jet edge becomes visible, 
the light-curve power-law decay steepens from that in equation (\ref{alpha}) to $F_\nu \propto t^{-p}$ 
(Rhoads 1999), \ie the light-curve index increases by
\begin{equation}
 \Da_{jet} = \frac{1}{4} \cdot  \left\{ \begin{array}{ll} 
           p+3 \;, & \nu < \nu_c \;\; \& \;\; s=0 \\ 
           p+2 \;, & \nu_c < \nu \;\; \& \;\; s=0,2 \\ 
           p+1 \;, & \nu < \nu_c  \;\; \& \;\; s=2  \end{array} \right.  \;.
\label{J}
\end{equation}

\begin{figure*}
\vspace*{5mm}
\centerline{\psfig{figure=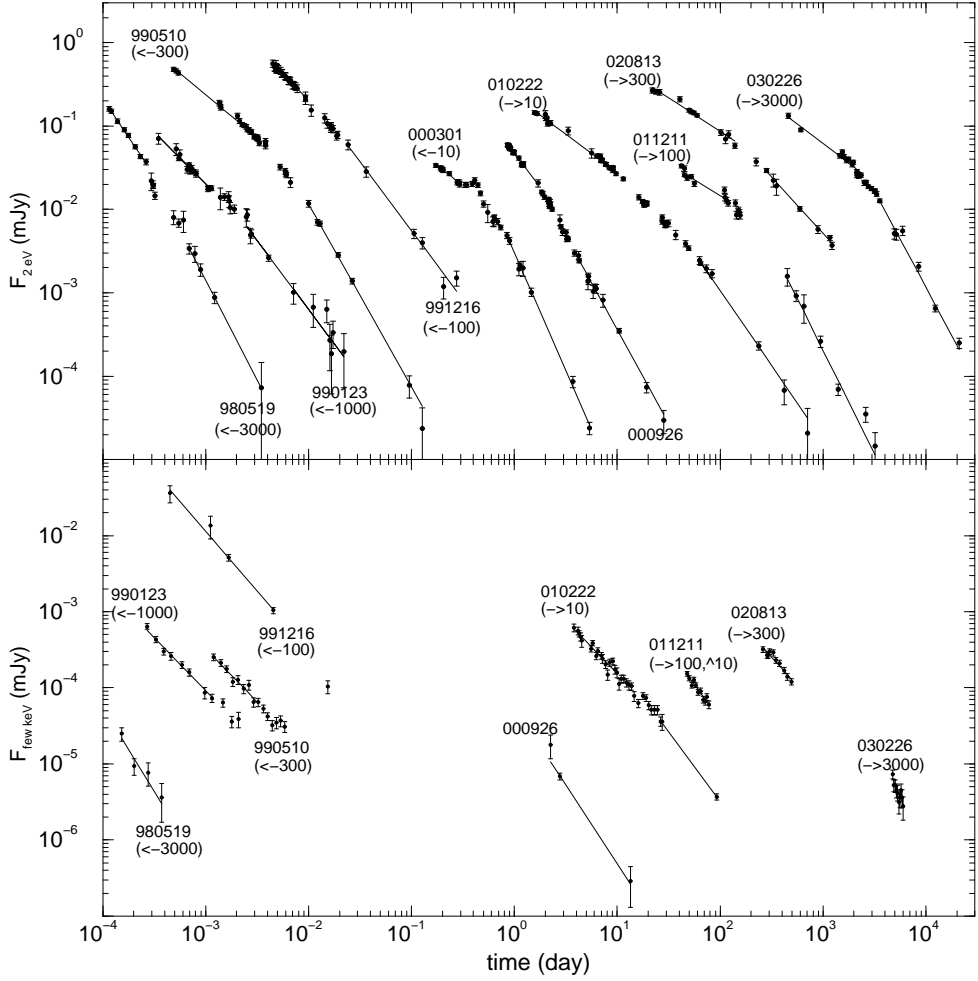,width=13cm}}
\caption{ Optical (upper panel) and $X$-ray (lower panel) light-curves for 10 GRB afterglows
   with breaks and power-law fits to the pre- and post-break emission. In most cases, the decay 
   indices given in Table 1 are averages of the fitting indices over a few bands, but only one 
   band ($R$) is shown here. The $X$-ray light-curves were calculated from the available wide 
   band (0.2--10 keV or 2-10 keV, usually) fluxes and SED slopes. 
   For clarity, light-curves have been shifted horizontally (to the left for "$\leftarrow$",
   to the right for "$\rightarrow$") by the indicated factors. For each afterglow, the shifting 
   factor is the same for both the optical and $X$-ray light-curves. 
   The data shown here and the optical measurements of Fig. 3--11 are taken from :
   980519 -- Halpern \etal (1999), Vrba \etal (2000), Jaunsen \etal (2001), Nicastro \etal (1999); 
   990123 -- Galama \etal (1999), Kulkarni \etal (1999), Piro (2000);
   990510 -- Harrison \etal (1999), Stanek \etal (1999), Kuulkers \etal (2000);
   991216 -- Garnavich \etal (2000), Halpern \etal (2000), Corbet \& Smith (1999), Piro \etal (1999), 
                Takeshima \etal (1999); 
   000301 -- Jensen \etal (2001), Rhoads \& Fruchter (2001);
   000926 -- Fynbo \etal (2001), Harrison \etal (2001), Price \etal (2001), Piro \etal (2001);
   010222 -- Masetti \etal (2001), Stanek \etal (2001), in't Zand \etal (2001);
   011211 -- Holland \etal (2002), Jakobsson \etal (2003), Borozdin \& Trudolyubov (2003);
   020813 -- Covino \etal (2003), Gorosabel \etal (2004), Butler \etal (2003);
   030226 -- Klose \etal (2004), Pandey \etal (2004).  }
\end{figure*}

\section{Model Selection Criteria}
\label{criteria}

 To summarize, in the framework of the above models, afterglows light-curves should exhibit
a steepening from $F_\nu \propto t^{-\alpha_1}$ to $F_\nu \propto t^{-\alpha_2}$, with: 
\begin{equation}
 {\rm Energy \; Injection \; (EI)} :\left\{ \begin{array}{l}  \alpha_1 = \alpha - \Da_{ei} \\
                            \alpha_2 = \alpha  \end{array} \right. \;,
\label{break1}
\end{equation}
\begin{equation}
 {\rm Structured \; Outflow \; (SO)} :\left\{ \begin{array}{l}  \alpha_1 = \alpha \\
                  \alpha_2 = \alpha + \Da_{so} \end{array} \right. \;,
\end{equation}
\begin{equation}
 {\rm Jet \; in \; Outflow \; (J/O)} :\left\{ \begin{array}{l}  \alpha_1 = \alpha \\
                        \alpha_2 = \alpha +  \Da_{j/o} \end{array} \right. \;,
\end{equation}
\begin{equation}
 {\rm Jet \; (J)} :\left\{ \begin{array}{l}  \alpha_1 = \alpha \\
                        \alpha_2 = \alpha + \Da_{jet}  \end{array} \right. \;,
\label{break2}
\end{equation}
where $\alpha$ is that given in equation (\ref{alpha}). 

 Equation (\ref{beta}) for the intrinsic SED slope $\beta_o$, equation (\ref{betaobs}) for the
observed slope $\beta$, and equations (\ref{break1})--(\ref{break2}), provide a simple criterion 
to determine from the observables $\alpha$ and $\beta$ which case is at work, for a given model
and afterglow:
\begin{equation} \left. \hleft \begin{array}{llll} 
  i) & \hleft 3\beta - 2 \alpha_{1,2} < -1       & \rightarrow &{\rm model \; does \; not \; work} \\
 ii) & \hleft 3\beta - 2 \alpha_{1,2} \in (-1,0) & \rightarrow & \nu_o < \nu_c \;\&\; s=2 \\
iii) & \hleft 3\beta - 2 \alpha_{1,2} \in (0,1)  & \rightarrow & \nu_o < \nu_c \;\&\; s=0,2 \\
 iv) & \hleft 3\beta - 2 \alpha_{1,2} > 1        & \rightarrow & {\rm no \;\; restriction}
   \end{array} \right. \;,
\label{ba}
\end{equation}
where $\nu_o \sim 3\cdot 10^{14}$ Hz. In the above criterion, the pre-break index $\alpha_1$ 
applies to the SO, J/O and J models, and the post-break index $\alpha_2$ is to be used for the
EI model. The first line of equation (\ref{ba}) states that even the largest possible intrinsic 
SED slope, $\beta_o = \beta$, is too small to be consistent with the steepness of the post-break 
light-curve decay. The following lines specify which cases are allowed; in each case the difference 
$\beta - \beta_o$ giving the host extinction $A_V$.

 Once the working cases for the EI model are identified, the break magnitude $\Da$ 
can be used to determine the exponent $e$ of the injection law. From equations (\ref{alpha}) and
(\ref{EI}):
\begin{equation}
 e = 3\,\Da \; \cdot \left\{ \begin{array}{ll} 
           (\alpha_2+3)^{-1} \;, & \nu_o < \nu_c \;\; \& \;\; s=0 \\ 
           (\alpha_2+2)^{-1} \;, & \nu_c < \nu_o \;\; \& \;\; s=0,2 \\ 
           (\alpha_2+1)^{-1} \;, & \nu_o < \nu_c  \;\; \& \;\; s=2  \end{array} \right.  \;.
\label{ee}
\end{equation}
The only restriction here is $e < 3$ for a homogeneous medium and $e < 1$ for a wind-like medium, 
otherwise the dynamics of the refreshed shocks would be inconsistent with the assumed injection 
law $E \propto t^e$.
 
 Taking the break magnitude given in equation (\ref{J}) as an upper limit for the $\Da$
allowed by the SO, J/O and J models, and keeping in mind that larger breaks could result from
structured outflows if the observer lies outside the core opening, one reaches the following
criterion for identifying the viable cases for structured outflows and jets:
\begin{equation} \left. \hleft \begin{array}{llll} 
  i) & \hleft  3\,\Da - \a1 > 3       & \hleft \rightarrow \hleft & {\rm observer \; outside \; core } \\
 ii) & \hleft  3\,\Da - \a1 \in (2,3) & \hleft \rightarrow \hleft & \nu_o < \nu_c \;\&\; s=0 \\
iii) & \hleft  3\,\Da - \a1 \in (1,2) & \hleft \rightarrow \hleft & s=0 \;{\rm or}\; \nu_c < \nu_o \;\&\; s=2 \\
 iv) & \hleft  3\,\Da - \a1 < 1       & \hleft \rightarrow \hleft & {\rm no \;\; restriction }
         \end{array} \right. \;. 
\label{jj}
\end{equation}
For the SO model, if the observer is located within the opening of the uniform core, the measured 
$\Da$ determines the exponent $q$ of the angular distribution of the ejecta kinetic energy 
(\eq [\ref{q}]). From equations (\ref{alpha}), (\ref{SO}) and (\ref{qt}), one obtains
\begin{equation}
  q = \Da \; \cdot \left\{ \begin{array}{lll}
       8/(\alpha_2+3)       &,& \nu_o < \nu_c \;\&\; s=0 \\ 
       8/(\alpha_2+2)       &,& \nu_c < \nu_o \;\&\; s=0 \\  
      12/(3\alpha_2-2\a1+1) &,& \nu_o < \nu_c \;\&\; s=2 \\
      12/(3\alpha_2-2\a1+2) &,& \nu_c < \nu_o \;\&\; s=2 \end{array} \right.  \;.
\label{qq}
\end{equation}
If the direction toward the observer is outside the outflow core, then the criterion given in equation
(\ref{jj}) does not hold. From the maximum post-break decay index allowed by the SO model
\begin{equation}
 \alpha_2 =  \frac{1}{4} \cdot \left\{ \begin{array}{lll}
                     3p   & \hLeft , & \hleft (s=0, \nu_o < \nu_c) \;\; {\rm or} \;\; (s=2, \nu_c < \nu_o) \\ 
                     3p+1 & \hLeft , & \hleft (s=0, \nu_c < \nu_o) \;\; {\rm or} \;\; (s=2, \nu_o < \nu_c)  
        \end{array} \right. 
\end{equation}
and from $\beta > \beta_o$, the criterion given in equation (\ref{ba}) is replaced by
\begin{equation} \left. \hleft \begin{array}{llll} 
  i) & \hLeft 3\beta - 2 \alpha_2 < -2          & \hleft \rightarrow & \hleft {\rm SO \; model \; does \; not \; work} \\
 ii) & \hLeft 3\beta - 2 \alpha_2 \in (-2,-1)   & \hleft \rightarrow & \hleft \nu_o < \nu_c \;\&\; s=2 \\
iii) & \hLeft 3\beta - 2 \alpha_2 \in (-1,-0.5) & \hleft \rightarrow & \hleft \nu_o < \nu_c \;\&\; s=0,2 \\
 iv) & \hLeft 3\beta - 2 \alpha_2 \in (-1/2,0)  & \hleft \rightarrow & \hleft s=0\; {\rm or}\; \nu_o < \nu_c \;\&\; s=2 \\
 iv) & \hLeft 3\beta - 2 \alpha_2 > 0           & \hleft \rightarrow & \hleft {\rm no \;\; restriction}
   \end{array} \right. \;.
\label{ba1}
\end{equation}

\section{Observations}
\label{observations}

\begin{table*}
\caption{ Light-curve decay indices and slopes of the optical and $X$-ray emission of 10 GRB
           afterglows with optical light-curve breaks }
\begin{tabular}{c|cc|ccc|ccc|c}
  \hline \hline
  GRB     &$t_1/t_b$&$t_2/t_b$& $\alpha_1$    & $\alpha_2$    & $\beta$        &  $\delta_1$   &  $\delta_2$          &   $\beta_x$       &  $\beta_{ox}$   \\
          &   (1)   &   (2)   &     (3)       &     (4)       &      (5)       &    (6)        &     (7)              &     (8)           &     (9)         \\
  \hline
  980519  &  0.3    &    3    & $1.87\pm0.11$ & $2.43\pm0.35$ & $1.07\pm0.12$  & $2.23\pm0.91$ &                      & $1.8\pm0.6^{(a)}$ &  $1.01\pm0.03$  \\
  990123  &  0.1    &$\siml10$& $1.17\pm0.05$ & $1.65\pm0.20$ & $0.63\pm0.05$  & $1.44\pm0.13$ &                      &                   &  $0.67\pm0.02$  \\
  990510  &  0.1    &   20    & $0.92\pm0.04$ & $2.21\pm0.09$ & $0.60\pm0.10$  & $1.40\pm0.19$ &                      &$1.03\pm0.08^{(b)}$&  $0.88\pm0.02$  \\
  991216  &  0.25   &$\sim10$ & $1.25\pm0.16$ & $1.65\pm0.12$ & $0.57\pm0.08$  & $1.57\pm0.14$ &                      &                   &  $0.84\pm0.02$  \\
  000301  &  0.3    &   10    & $1.00\pm0.13$ & $2.83\pm0.12$ & $0.99\pm0.05$  &               &                      &                   &                 \\
  000926  &  0.4    &   10    & $1.57\pm0.11$ & $2.21\pm0.06$ & $1.31\pm0.06$  &               & $2.06\pm0.22$        &$0.96\pm0.13^{(c)}$&  $0.94\pm0.03$  \\
  010222  &  0.1    &$\siml20$& $0.90\pm0.02$ & $1.78\pm0.08$ & $1.05\pm0.09$  & $1.39\pm0.15$ & $1.83\pm0.20\dagger$ &$0.97\pm0.05^{(d)}$&  $0.64\pm0.01$  \\
  011211  &  0.1    &   10    & $0.79\pm0.07$ & $2.49\pm0.24$ & $0.71\pm0.07$  & $1.73\pm0.17$ &                      &$1.18\pm0.10^{(e)}$&  $1.06\pm0.01$  \\
  020813  &  0.1    &    6    & $0.80\pm0.04$ & $1.34\pm0.05$ & $1.13\pm0.07$  &               & $1.52\pm0.13$        &$0.85\pm0.04^{(f)}$&  $0.63\pm0.04$  \\
  030226  &  0.2    &    8    & $0.89\pm0.03$ & $2.37\pm0.06$ & $0.93\pm0.05$  &               & $3.4\pm1.5$          &$1.04\pm0.20^{(g)}$&  $1.02\pm0.05$  \\
  \hline \hline
\end{tabular}
\begin{minipage}{170mm}
 {\bf (1)} ratio of epochs of first optical measurement and light-curve break ;
 {\bf (2)} ratio of epochs of break and last accurate measurement of optical transient brightness ;
 {\bf (3)} pre-break optical light-curve index ($F_\nu \propto t^{-\alpha_1}$) ;
 {\bf (4)} post-break optical light-curve index  ;
 {\bf (5)} observed slope of optical SED ($F_\nu \propto \nu^{-\beta}$) ;
 {\bf (6)} pre-break $X$-ray light-curve index  ;
 {\bf (7)} post-break $X$-ray light-curve index ($\dagger$ relying on a single, 10 day measurement) ;
 {\bf (8)} slope of $X$-ray SED (Refs: $^{(a)}$ Nicastro \etal (1999), $^{(b)}$ Kuulkers \etal (2000), 
           $^{(c)}$ Harrison \etal (2001) and Piro \etal (2001), $^{(d)}$ in't Zand \etal (2001), 
           $^{(e)}$ Reeves \etal (2003), $^{(f)}$ Butler \etal (2003), $^{(g)}$ Klose \etal (2004) ;
 {\bf (9)} optical to $X$-ray SED slope (\eq [\ref{box}]) 
\end{minipage}
\end{table*}

 Before proceeding with testing the models presented in \S\ref{models}, we determine the relevant 
observables for a sample of 10 GRB afterglows whose optical light-curves exhibit a break. The pre- 
and post-break decay indices, $\alpha_1$ and $\alpha_2$, of the optical light-curves were obtained 
through power-law fits to the afterglow light-curve at times before and after the break, excluding 
measurements close to the break time. Therefore, the fitted data stretch a time range somewhat shorter 
than given in columns 2 and 3 of Table 1. Such fits are shown in only one band, for each afterglow, 
in the upper panel of Fig. 1. With the exception of 000301, all pre-break indices $\alpha_1$ are 
averages of values determined for 3 or 4 optical bands ($I$, $R$, $V$, and $B$). However, for half 
of the afterglows, sufficient data to determine the post-break index exist in only one band (usually 
the $R$ band). The lower panel of Fig. 1 shows the power-law fits to the $X$-ray emission, used
to determine the pre- or post-break decay indices $\delta_1$ and $\delta_2$. With the exception of
010222, the existing data allows the determination of only one of the indices; for 010222, the
existence of a break relies on single, late-time measurement.


 Fig. 2 shows the distribution of the break magnitude $\Da$. With the current sample of ten 
afterglows with breaks, there is only a vague suggestion for a bimodal distribution, with a gap 
at about $\Da = 1$. Such a bimodality, if real, would not be a natural feature of the Energy 
Injection model, where any value of $\Da$ is (a priori) equally likely, but could arise from 
Structured Outflows, where the smaller break magnitudes are due to the ejecta angular structure, 
while the larger values are for spreading jets. The gap at $\Da = 1$ would then point to a 
homogeneous medium and cooling frequency above the optical.

 Figs. 3--12 show the slope $\beta$ of the optical SED for each afterglow, at a few epochs.
With the exception of 011211, these epochs bracket the light-curve break time, extending over 
approximately one decade in time. We do not find any compelling evidence for a spectral evolution 
in any afterglow, though, given the uncertainty of the instantaneous SED slopes, the crossing of
a slowly evolving, shallow, and smooth spectral break (such as the cooling break) cannot be ruled 
out.  For this reason, we do not consider in this work light-curve breaks arising from the passage 
of a spectral break through the observing band. Such a scenario would require a break in the electron 
energy distribution which is stronger than that due to cooling or which evolves faster. 
The evolution of the cooling break ($\nu_c \propto t^{-1/2}$ for $s=0$ and $\nu_c \propto t^{1/2}$ 
for $s=2$, assuming synchrotron-dominated radiative losses) and the steepening $\Delta \beta = 1/2$ 
of the SED across it imply that the passage of $\nu_c$ through the optical domain would yield only 
a modest break magnitude: $\Da = 1/4$. 
 
 The optical slopes $\beta$ reported in Table 1 are averages of the instantaneous values shown in
Fig. 3--12. Besides reducing the uncertainty, averaging over a few epochs is also a safeguard 
against short-term colour fluctuations. Optical fluxes were corrected for Galactic extinction but 
not for a possible host reddening. While the curvature of the optical SED due to the differential dust 
extinction produced by $A_V \sim 0.1$ mag is usually masked by the errors of measuring the afterglow 
I, R, V, B band fluxes, near-infrared ($K$, $H$, $J$ bands) measurements widen enough the frequency 
baseline that they allow the measurement of that curvature, the determination of $A_V$ and of 
$\beta_o$ (\eg Fynbo \etal 2001, Jensen \etal 2001, Jakobsson \etal 2003). Aside for the uncertainty 
in the fitted $\beta_o$ introduced by assuming a reddening curve, a second problem with inferring 
$\beta_o$ from the observed curvature of the afterglow optical SED is that the intrinsic 
spectrum may be curved if the cooling frequency $\nu_c$ is within the NIR-optical domain, a 
coincidence which may not be that rare.  In such a case, the host extinction will be overestimated.

 An other way of inferring $A_V$ and $\beta_o$ is to compare the $X$-ray flux with the extrapolation
to the $X$-ray domain of the optical SED. If the measured $X$-ray flux lies on that extrapolation,
then it is tempting to conclude that $A_V \simeq 0$, while an $X$-ray excess may be seen as the
signature of optical host extinction. In the latter case, from the optical-to-$X$-ray flux ratio 
and the optical and $X$-ray SED slopes ($\beta_o$ and $\beta_x$), it is possible to determine the 
$\nu_c$, $\beta_o$, and $A_V$ whenever $\beta_x$ is known sufficiently well (\eg Galama \& Wijers 2001). 
Otherwise, if $\nu_c$ is assumed not to be in between the optical and $X$-ray domains, matching the \
extrapolation of the dereddened optical continuum with the $X$-ray flux could lead to an underestimation 
of $A_V$ and overestimation of $\beta_o$. In the worst case -- $\nu_c \simg \nu_o$  -- the $X$-ray flux 
would lie on the extrapolation of the optical SED for $A_V = 0.5/(z+1)$. 

 Perhaps the largest uncertainty in determining $\nu_c$, $\beta_o$, and $A_V$ as described above arises 
from the possible contribution of inverse Compton scatterings to the $X$-ray afterglow emission, which
alters the optical-to-$X$-ray flux ratio and the $X$-ray SED slope. If inverse Compton scatterings are 
ignored, it could lead to an overestimated $A_V$ and underestimated $\beta_o$. 
For this reason, we do not determine $\beta_o$ and $A_V$ directly from observations, but infer them 
from the observed SED slopes and light-curve decay indices, using the relationship between them that 
is expected for each model. For further use, we also measure and list in Table 1 the observed 
optical-to-$X$-ray slope, defined by:
\begin{equation}
 \beta_{ox} = - \frac{\ln F_x/F_o}{\ln \nu_x/\nu_o} \;,
\label{box}
\end{equation}
where the subscripts "x" and "o" denote $X$-ray and optical.
As we shall see, the inferred host extinction $A_V$ is usually less than 0.2 mag, \ie the intrinsic 
optical-to-$X$-ray slope is only slightly larger than the observed one.

 The $1\sigma$ dispersion of the indices $\alpha_1$, $\alpha_2$, $\beta$, $\delta_1$, $\delta_2$,
and $\beta_{ox}$ given in Table 1 is for the joint variation of both the normalization constant
and the index, \ie it is the width of the projection of the $\Delta \chi^2 = 2.30$ contour on
the $\alpha$-axis.

\begin{figure}
\centerline{\psfig{figure=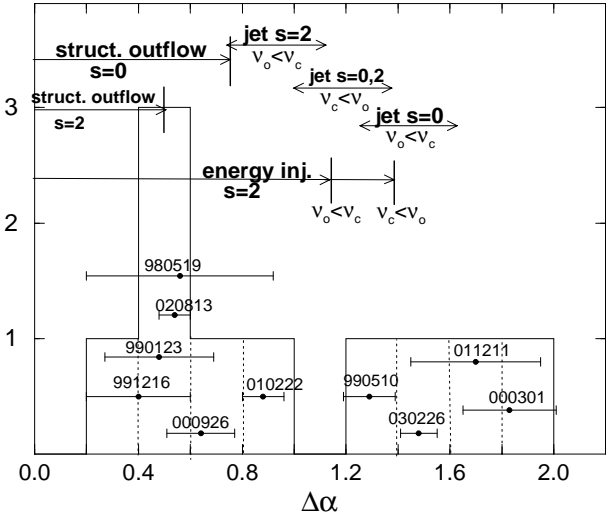,width=8cm}}
\caption{ Histogram of the break magnitude $\Da$, the increase of the afterglow power-law decay 
  index, for the optical light-curves of the 10 afterglows considered in this work. At the top 
  are indicated the values expected theoretically for structured outflows, jets, and refreshed 
  shocks (energy injection model), interacting with homogeneous media ($s=0$) and free winds 
  ($s=2$), for a cooling frequency ($\nu_c$) below or above the optical domain ($\nu_o$).
  Vertical bars indicate upper limits. For structured outflows they correspond to a direction 
  toward the observer within the opening of the uniform outflow core; otherwise $\Da$ can be larger.
  For energy injection and $s=2$, the upper limits are for an electron energy distribution 
  (\eq [\ref{dNde}]) with $p=3.5$, while the $s=0$ medium covers the entire $\Da$ range shown. 
  For jets, the ranges of $\Da$ shown correspond to $p \in (2.0,3.5)$.}
\end{figure}

\begin{figure}
\centerline{\psfig{figure=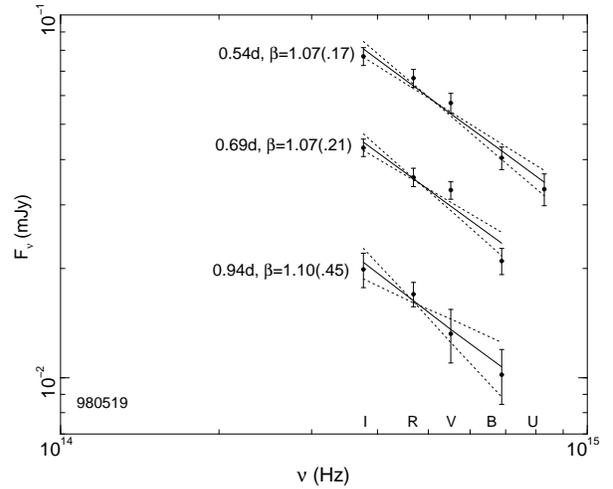,width=8cm}}
\caption{ Optical slope $\beta$ for the afterglow 980519, at various epochs (in days). 
  $1\sigma$ uncertainties of the slope are given in parentheses.
  Fluxes are corrected for Galactic extinction of $E(B-V)=0.27$. Solid lines are the best power-law 
  fits; the dotted lines show the power-laws of slopes $1\sigma$ above and below that of the best fit.}
\end{figure}

\begin{figure}
\centerline{\psfig{figure=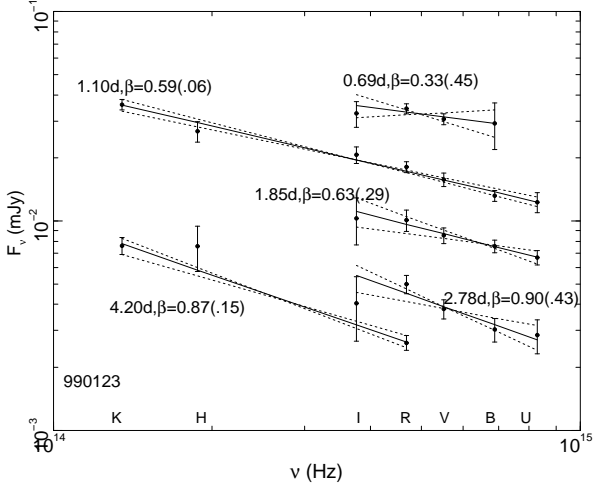,width=8cm}}
\caption{Afterglow 990123, $E(B-V)=0.02$}
\end{figure}

\begin{figure}
\centerline{\psfig{figure=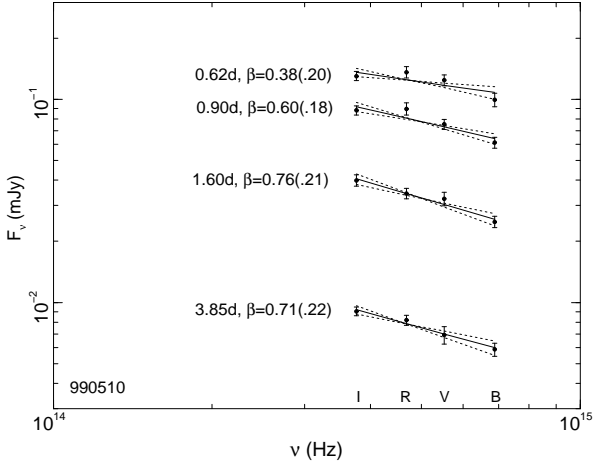,width=8cm}}
\caption{Afterglow 990510, $E(B-V)=0.20$}
\end{figure}

\begin{figure}
\centerline{\psfig{figure=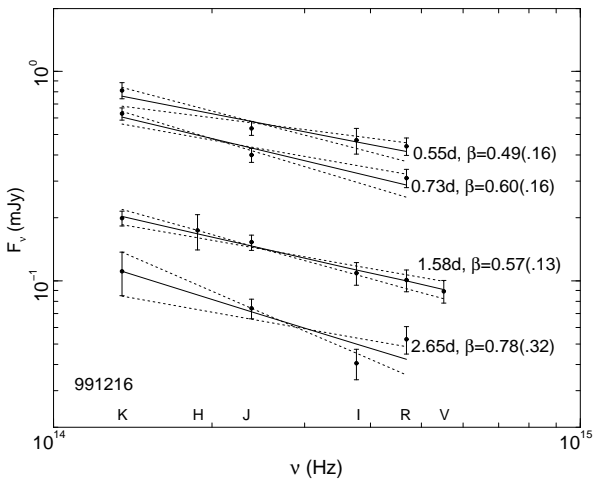,width=8cm}}
\caption{Afterglow 991216, $E(B-V)=0.63$}
\end{figure}

\begin{figure}
\centerline{\psfig{figure=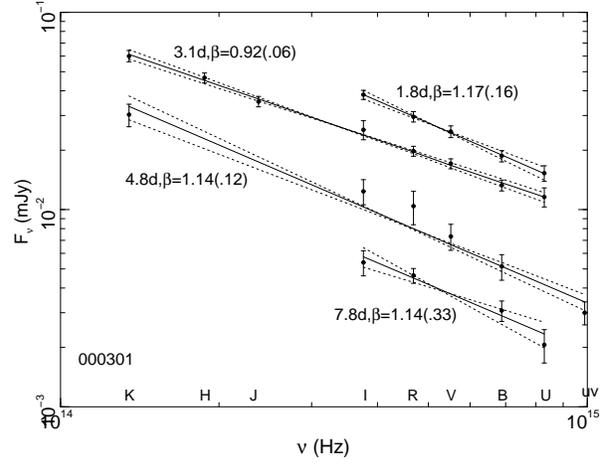,width=8cm}}
\caption{Afterglow 000301, $E(B-V)=0.05$}
\end{figure}

\begin{figure}
\centerline{\psfig{figure=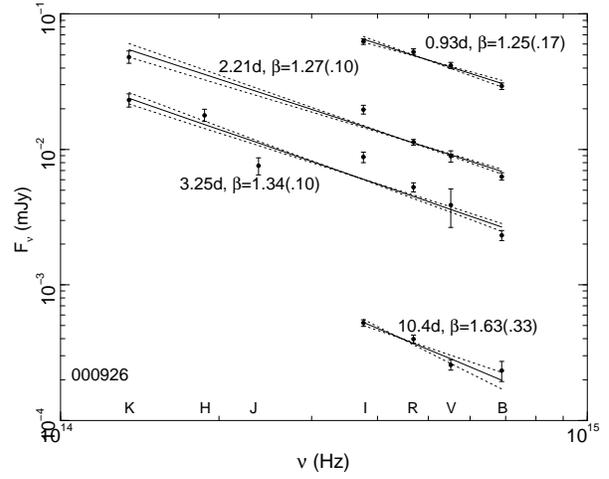,width=8cm}}
\caption{Afterglow 000926, $E(B-V)=0.02$}
\end{figure}

\begin{figure}
\centerline{\psfig{figure=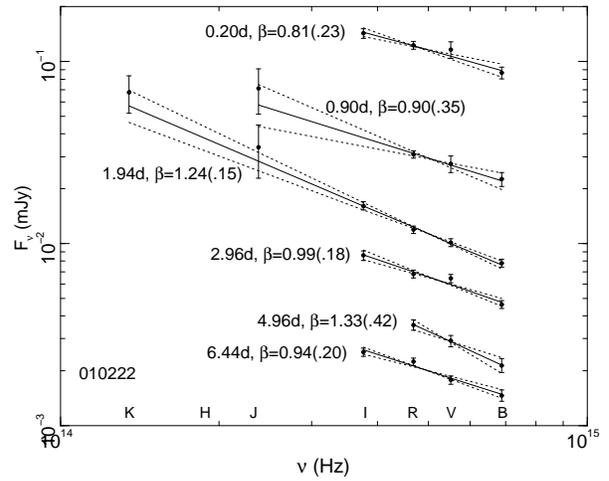,width=8cm}}
\caption{Afterglow 010222, $E(B-V)=0.02$}
\end{figure}

\begin{figure}
\centerline{\psfig{figure=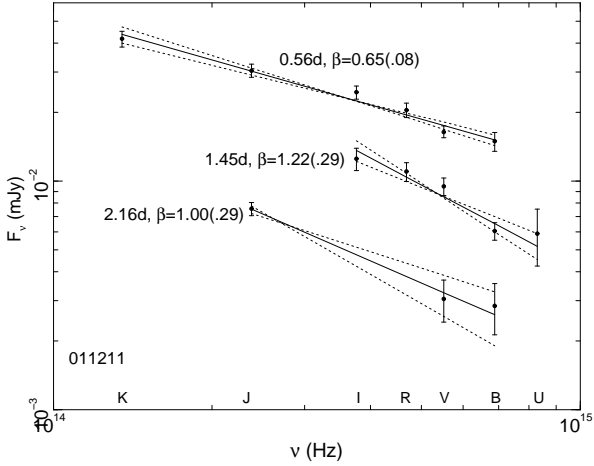,width=8cm}}
\caption{Afterglow 011211, $E(B-V)=0.04$}
\end{figure}

\begin{figure}
\centerline{\psfig{figure=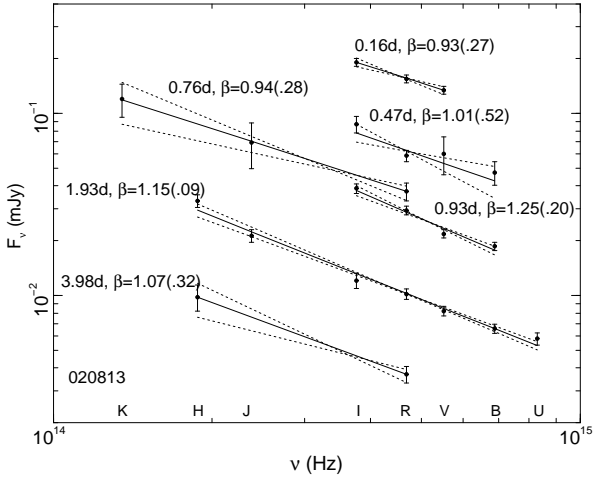,width=8cm}}
\caption{Afterglow 020813, $E(B-V)=0.10$}
\end{figure}

\begin{figure}
\centerline{\psfig{figure=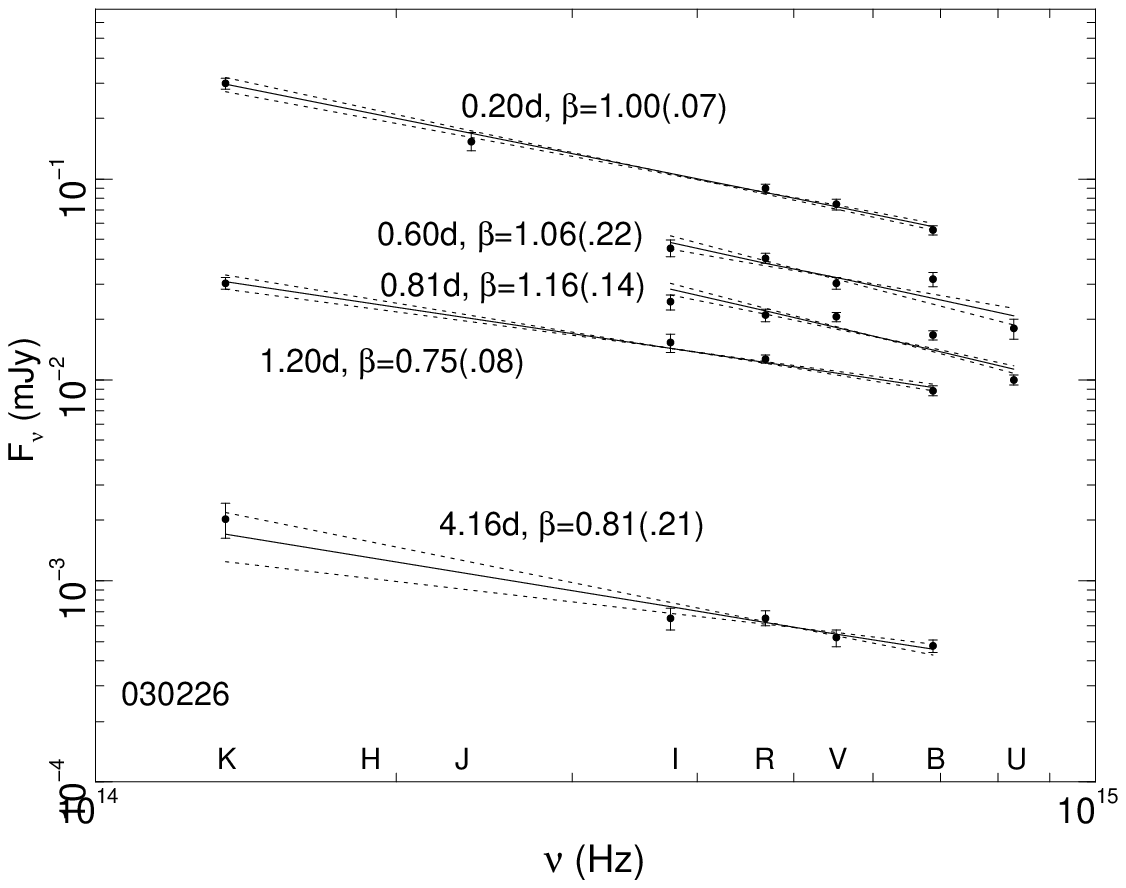,width=8cm}}
\caption{Afterglow 030226, $E(B-V)=0.02$}
\end{figure}

\section{Model Tests and Results}
\label{tests}

 We use the criteria given in equations (\ref{ba}) and (\ref{jj}) to select from the four models 
for light-curve breaks -- Energy Injection, Structured Outflow, (non-spreading) Jet in Outflow, 
and (spreading) Jet -- those which can accommodate the properties of the optical emission 
($\alpha_1$, $\alpha_2$, and $\beta$ of Table 1) of the GRB afterglows shown in Fig. 1. 
To take into account the uncertainty of the measured light-curve indices and SED slopes,
we set the following requirements for acceptable models.

 The observed SED slope, $\beta$, should be larger or equal than the intrinsic one, $\beta_o$. 
Using equations (\ref{alpha}) and (\ref{beta}), we calculate the model SED slope $\beta_{model}$ 
from the pre-break light-curve index $\alpha_1$ for models SO, J/O and J, and from the post-break 
index $\alpha_2$ for model EI, and require that 
\begin{equation}
 {\rm constraint\; 1\;:\;} \beta > \beta_{model}  - 3\; \sigma (\beta - \beta_{model}) \;,
 \label{btest}
\end{equation}
where $\sigma$ is the uncertainty of the slope difference. 

 For the SO model, the measured break magnitude $\Da$ should be smaller than the maximum 
allowed $\Da_{model}$ given in equation (\eq [\ref{JO}]), which is equivalent to requiring
that $q < \tilde{q}$ (\eq [\ref{qt}]):
\begin{equation}
 {\rm constraint\; 2a\;:\;} \Da < \Da_{model} + 3\; \sigma (\Da - \Da_{model}) \;,
\end{equation}
Given that, for jets, the transition to the post-break decay takes almost a decade in time 
for a homogeneous medium and two decades for a wind-like medium (Kumar \& Panaitescu 2000) 
and that, for most afterglows, the post-break light-curve is monitored for about one decade 
in time, it is possible that the measured decay index $\alpha_2$ underestimates its asymptotic 
value. For this reason, we impose the following constraint for the J/O and J models:
\begin{equation}
 {\rm constraint\; 2b\;:\;} \Da_{model} - 5\; \sigma_\Delta < \Da < \Da_{model} + 3\; \sigma_\Delta \;,
\end{equation}
where $\sigma_\Delta \equiv \sigma (\Da - \Da_{model})$ and $\Da_{model}$ is that given by 
equations  (\eq [\ref{JO}]) and (\eq [\ref{J}]), respectively. \\
For the EI model, the observed $\Da$ determines the injection parameter $e$ (\eq [\ref{e}]), 
for which the only constraint is $e < 3-s$. 

 A third condition is that the exponent $p$ of the power-law electron distribution with energy 
(\eq [\ref{dNde}]) required by the observed $\alpha_1$ for the SO, J/O, and J models, and by the 
$\alpha_2$ for the EI model should be larger than 2
\begin{equation}
 {\rm constraint\; 3\;:\;} p + 3\,\sigma(p) > 2 \;.
\end{equation}
This condition is set to avoid an electron energy distribution with a cut-off or bend at higher 
energies (otherwise the total electron energy diverges). Such an electron distribution would provide 
an explanation for why the radio decay of some afterglows is significantly slower than at optical 
frequencies (Li \& Chevalier 2001, Panaitescu \& Kumar 2002), but here we choose to ignore this 
somewhat ad-hoc scenario.

\begin{table*}
 \caption{ Possible models for the 10 GRB afterglows of table 1 } 
\begin{tabular}{lllllllll}  
  \hline \hline
  GRB     &   Model          & $s$ & $e$ or $q$ &  $p$     & $\beta_o$  & $A_V$ &   Note                             &  Requirement from $X$-ray            \\
          &    (1)           & (2) &    (3)     &  (4)     &   (5)      &  (6)  &   (7)                              &  (8)                                 \\
  \hline 
  980519  &  [EI]            &  0  & $e\sim0.3$ &$\simg3.7$& $\simg1.4$ &   0   & $\beta < \beta_o$ at $2.1\,\sigma$ &  IC in $X$-ray                       \\ 
          &   EI             &  2  & $e\sim0.5$ &$\simg3.1$& $\simg1.0$ &   0   &                                    &  inconclusive                        \\
          & (SO),(J/O),[J]   &  0  & $q\sim0.8$ &  3.5     &   1.25     &   0   & $\beta < \beta_o$ at $1.2\,\sigma$ &  $\nu_x < \nu_c$                     \\
          &  SO,J/O,J        &  2  & $q\sim1.4$ &  2.8     &   0.92     &  0.1  &                                    &  $\nu_c < \nu_x$                     \\ 
  \hline
  990123  &   EI             &  2  & $e\sim0.5$ &  2.5     &   0.76     &   0   &                                    &  $\nu_x < \nu_c$                     \\ 
          & [SO],[J/O],\{J\} &  0  & $q\sim0.8$ &  2.6     &   0.78     &   0   & $\beta < \beta_o$ at $2.1\,\sigma$ &  $\nu_c < \nu_x$                     \\
          &  {\bf SO,J/O,J}  &  2  & $q\sim1.6$ &  1.9     &   0.45     &  0.1  &                                    &  $\delta_1-\a1>-1/4$ ICW $\beta_{ox}>\beta_o$  \\ 
  \hline
  990510  &   J              &  0  &            &  2.2     &   0.61     &   0   &                                    &  $\nu_c < \nu_x$                     \\
  \hline
  991216  &  [EI]            &  2  & $e\sim0.5$ &  2.5     &   0.77     &   0   & $\beta < \beta_o$ at $2.4\,\sigma$ &  $\nu_x < \nu_c$                     \\
          & [SO],[J/O],\{J\} &  0  & $q\sim0.7$ &  2.7     &   0.84     &   0   & $\beta < \beta_o$ at $2.1\,\sigma$ &  $\nu_c < \nu_x$                     \\
          & {\bf SO,J/O,(J)} &  2  & $q\sim1.4$ &  2.0     &   0.50     &   0   & J: $3\,\Da<\a1+1$ at $1.5\,\sigma$ &  $\delta_1-\a1>-1/4$ ICW $\beta_{ox}>\beta_o$  \\ 
  \hline
  000301  & [J]              &  0  &            &  2.3     &0.67$^{(a)}$&  0.1  & $3\,\Da > \a1+3$ at $2.4\,\sigma$  &  no $X$-ray data                     \\ 
  \hline
  000926  & [EI]             &  0  & $e=0.4$    &  3.9     &1.47$^{(b)}$&   0   & $\beta <\beta_o$ at $2.2\,\sigma$  &  IC in $X$-ray                       \\
          &  EI              &  2  & $e=0.6$    &  3.3     &1.14$^{(b)}$&  0.05 &                                    &  IC in $X$-ray                       \\
          &  SO,J/O,\{J\}    &  0  & $q=1.0$    &  3.1     &1.04$^{(b)}$&  0.1  & J: $3\,\Da<\a1+3$ at $4.8\,\sigma$ &  $\nu_x < \nu_c$                     \\
          &  SO,J/O,(J)      &  2  & $q=1.3$    &  2.4     &0.72$^{(b)}$&  0.15 & J: $3\,\Da<\a1+1$ at $1.2\,\sigma$ &  $\nu_c < \nu_x$                     \\
          & SO$\dagger$,J/O$\dagger$,[J]$\dagger$
                             & 0,2 & $q=1.2$    &  2.8     &1.38$^{(b)}$&   0   & J: $3\,\Da<\a1+1$ at $3.0\,\sigma$ &  IC in $X$-ray                       \\
  \hline
  010222  & (EI)             &  0  & $e=0.6$    &  3.4     &   1.19     &   0   & $\beta <\beta_o$ at $1.1\,\sigma$  &  IC in $X$-ray                       \\ 
          &  EI              &  2  & $e=0.9$    &  2.7     &   0.85     &  0.1  &                                    &  IC in $X$-ray                       \\
          & (SO),(J/O),\{J\} &  0  & $q=1.3$    &  2.2     &   0.60     &  0.2  & SO,J/O: $\Da>3/4$ at $1.6\,\sigma$ &  $\nu_c < \nu_x$                     \\
  \hline
  011211  & {\bf (J)}        &  0  &            &  2.1     &0.53$^{(c)}$&  0.2  & $3\,\Da > \a1+3$ at $1.7\,\sigma$  &  $\delta_1 - \alpha_1 > 1/4$ at $3.8\,\sigma \Rightarrow$ IC ?        \\ 
  \hline
  020813  &  EI              &  0  & $e=0.4$    &  2.8     &   0.91     &  0.1  &                                    &  IC in $X$-ray                       \\ 
          &  EI              &  2  & $e=0.7$    &  2.2     &   0.57     &  0.25 &                                    &  $\nu_c \simeq \nu_x$                \\
          & (EI)$\dagger$    & 0,2 & $e=0.5$    &  2.5     &   1.24     &   0   & $\beta < \beta_o$ at $1.4\,\sigma$ &  IC in $X$-ray                       \\
          & SO,\{J/O\}       &  0  & $q=1.0$    &  2.1     &   0.53     &  0.30 & J/O: $\Da < 3/4$ at $3.0\,\sigma$  &  $\nu_c < \nu_x$                     \\
  \hline
  030226  & [J]              &  0  &            &  2.2     &   0.59     &  0.1  & $3\,\Da > \a1+3$ at $2.5\,\sigma$  &  $\nu_c < \nu_x$                     \\ 
  \hline \hline
\end{tabular} 
\begin{minipage}{180mm}
  {\bf (1)} EI = energy injection, SO = structured outflow, J/O = jet confined by outer outflow J = jet expanding sideways; for unbracketed models the constraints 
            1 and 2 are satisfied within $1\,\sigma$; for models shown in ( ) and [ ] parentheses, the requirement given in the column (7) is satisfied at the 
            $1\,\sigma-2\,\sigma$ and $2\,\sigma-3\,\sigma$ levels, respectively; \{ \} parentheses indicate J/O and J models for which the measured $\alpha_2 - 
            \alpha_1$ is $3\,\sigma - 5\,\sigma$ below the theoretically expected value ;
  {\bf (2)} $s=0$: uniform circumburst medium, $s=2$: wind-like medium ;
  {\bf (3)} parameter for energy injection (\eq [\ref{e}]) or outflow structure (\eqs [\ref{q}] and [\ref{qt}]), inferred from equations (\ref{ee}), and (\ref{qq}) ;
  {\bf (4)} power-law exponent of electron distribution with energy (\eq [\ref{dNde}]), resulting from equation (\ref{alpha}) (for 980519, the $\simg$ symbol is followed 
            by $p -\sigma_p$) ;
  {\bf (5)} slope of the intrinsic optical spectral energy distribution; for comparison, the following intrinsic afterglow SED slopes and host extinctions were 
            inferred from power-law fits to the optical SED and a host SMC-like reddening curve: 
            $^{(a)}$ $\beta_o = 0.57 \pm 0.02$ (at 3 days) and $A_V = 0.14 \pm 0.01$ (Jensen \etal 2001), 
            $^{(b)}$ $\beta_o = 1.00 \pm 0.18$ (at 1 day) and $A_V = 0.18 \pm 0.06$ (Fynbo \etal 2001), 
            $^{(c)}$ $\beta_o = 0.56 \pm 0.19$ (at 0.6 days) and $A_V = 0.08 \pm 0.08$ (Jakobsson \etal 2003) ; 
  {\bf (6)} dust extinction in host frame inferred from equations (\ref{beta}) and (\ref{betaobs}) ;
  {\bf (8)} the optical emission properties are incompatible with those at $X$-rays (constraints 4--6) for the models in bold face (ICW = "in contradiction with") .
  $\dagger$ the cooling frequency is below (redward of) the optical domain and the electron radiative cooling is assumed to be weaker than synchrotron 
            losses; for all other models, the cooling frequency is above (blueward of) the optical.
\end{minipage}
\end{table*}

 The models which satisfy the constraints 1, 2, and 3 above are given in Table 2. 
Further testing of the viable models is done using the $X$-ray light-curve decay 
index, $\delta$, the optical to $X$-ray slope, $\beta_{ox}$, and the $X$-ray SED slope, $\beta_x$.
If the optical and $X$-ray decay indices are equal, then the cooling frequency $\nu_c$ is not between 
the optical and $X$-ray domains (with the exception of the J model at $t > t_b$, in which case $\nu_c$ 
is constant in time). Consequently, the $X$-ray flux should lie on the extrapolation of the optical 
SED to $X$-rays and the slopes of the optical and $X$-ray continua should be the same:
\begin{equation}
 {\rm constraint\; 4\;:\;} \delta = \alpha \Rightarrow \beta_o = \beta_{ox} = \beta_x  \;.
\label{x1}
\end{equation}
If the optical and $X$-ray decay indices are different, then $\nu_c$ is between the two domains, 
in which case the $X$-ray flux lies below the extrapolation of the optical SED and the $X$-ray 
spectral slope is larger than that in the optical (\eq [\ref{beta}]):
\begin{equation}
 {\rm constraint\; 5\;:\;} \delta \neq \alpha \Rightarrow  
           \beta_o \leq \beta_{ox} \leq \beta_x  \;{\rm and}\; \beta_x = \beta_o + 1/2  \;.
\label{x2}
\end{equation}
Furthermore, in this case equation (\ref{alpha}) leads to 
\begin{equation}
  {\rm constraint\; 6a\;:\;} \delta - \alpha = \left\{ \begin{array}{lll} 
                 1/4 &,& s=0 \\ -1/4 &,&  s=2  \end{array} \right. 
\label{xo}
\end{equation}
for the EI model at $t > t_b$, the SO and J models at $t < t_b$, and the J/O model at all times.
For the EI model at $t < t_b$
\begin{equation}
  {\rm constraint\; 6b\;:\;} \delta - \alpha = \left\{ \begin{array}{lll} 
                  (1+e)/4 &,& s=0 \\ -(1+e)/4 &,& s=2  \end{array} \right.  \;, 
\end{equation}
while for the SO model at $t > t_b$
\begin{equation}
  {\rm constraint\; 6c\;:\;} \delta - \alpha = \left\{ \hleft \begin{array}{ll}
        (2-q)/(8-q) \;\;\;\; , & s=0 \\ -(2-q)/(8-2q) ,  & s=2  \end{array} \right.  \;.
\label{xq}
\end{equation}
Equations (\ref{xo}) -- (\ref{xq}) assume that the electron radiative cooling is dominated 
by synchrotron losses. If electrons cool mostly through inverse Compton scatterings, a situation which 
we ignore in this work, then $\delta - \alpha$ is smaller. 

 If the $X$-ray afterglow emission is mostly due to inverse Compton scatterings, then equations 
(\ref{x1}) -- (\ref{xq}) are not valid. Such a case is easily identifiable through that the $X$-ray 
emission is above the extrapolation of the optical SED 
\begin{displaymath}
 \beta_{ox} < \beta_o \Rightarrow {\rm Inverse\;\; Compton\;\; (IC)\;\;in\;\; X-ray} \;.
\end{displaymath}
However, a substantial contribution to the $X$-ray emission from inverse Compton scatterings cannot 
be excluded when $\beta_{ox} > \beta_o$ because the cooling frequency may lie below the $X$-ray
domain. 

 Whenever $\beta_{ox} \geq \beta_o$, we impose the conditions given in equations (\ref{x1}) -- 
(\ref{xq}) at the $3\,\sigma$ level (just as for the optical constraints). As shown in Table 2, 
only a few models are eliminated by the $X$-ray constraints.

\section{Conclusions}

 The first conclusion to be drawn from Table 2 is that an energy injection (the {\bf EI} model), 
mediated by some initially slower ejecta which catch up with the shock-front (the "refreshed shock" 
model), may be at work in many afterglows, although not in all the afterglows with breaks. For the 
same electron index $p$, a wind-like medium yields a faster light-curve decay than a homogeneous 
one, which makes the free wind medium more often compatible with the steepness of the post-break decay. 
The break magnitude $\Da$ determines the exponent of the assumed power-law energy injection law 
(\eq [\ref{e}]). For a homogeneous medium, we obtain $0.3 < e < 0.6$, while for a free wind, 
$0.5 < e < 0.9$, 
which imply that the total energy output until the break time ($t_b$) is larger than the post-burst
energy of the GRB ejecta by a factor of at least $(t_b/t_1)^e \simeq 1.5 - 4$ ($2-10$ for a free 
wind), where $t_1$ is the epoch of the first optical measurement (see 2nd column of Table 1). 
If the afterglow energy after the end of the energy injection process is close to that previously 
inferred by us from afterglow fits using the jet model (Panaitescu \& Kumar 2002), then the GRB
output is about 80 per cent of the ejecta initial energy. Such a high efficiency of the GRB mechanism 
is a serious problem (Kumar 1999) for the internal shock model.

 The resulting range for the $e$ parameter and equation (\ref{gamma}) imply that the distribution
of energy with Lorentz factor is $dE/d\Gamma \propto \Gamma^{-\gamma}$ with $\gamma \in (2,3)$ for
a homogeneous medium and $\gamma \geq 5$ for a free wind. The numerical calculations of Aloy \etal (2000) 
for the jet propagation through a collapsing massive star yield at break-out $dE/d\Gamma \propto 
e^{-0.7 \Gamma}$ for $\Gamma \in (5,30)$. A log-log fit to the $\log (dM/d\Gamma) - \Gamma$ plot 
shown in their Fig. 4 yields $dE/d\Gamma \propto \Gamma^{-5.3}$ for $\Gamma \in (5,15)$ and 
$dE/d\Gamma \propto \Gamma^{-11}$ for $\Gamma \in (10,30)$. Therefore the distribution of ejecta 
energy with Lorentz factor required by the EI model is compatible with that predicted by Aloy \etal 
(2000) only if the circumburst medium has a wind-like stratification.

 In contrast to the EI model, where the break magnitude $\Da$ is determined by the injection 
exponent $e$ and, hence, can have an arbitrary distribution, a structured outflow (extending 
this concept to the extreme case of a jet with sharp edges) can yield a bimodal distribution of 
$\Da$: $\Da \leq 3/4$ arises when the outflow envelope outshines the core (the {\bf SO} model), 
or when it is important only dynamically, by impeding the core's lateral spreading (the {\bf J/O} 
model), while afterglows with $\Da \geq 3/4$ arise from outflows with a sufficiently tenuous 
envelope that the core can expand sideways (the {\bf J} model). 
The current sample of 10 afterglows with light-curve breaks is too small to test the existence 
of a bimodal $\Da$ distribution. If the tentative gap at $\Da \sim 1$ shown in Fig. 2 is real, 
it will favour a homogeneous medium around GRBs.

 The second conclusion is that the SO model may be at work in more than half of the analyzed 
afterglows. In these cases, a $dE/d\Omega \propto \theta^{-1.0 \pm 0.3}$ angular distribution of 
the ejecta kinetic energy per solid angle is required for a homogeneous medium and  $dE/d\Omega 
\propto \theta^{-1.3 \pm 0.1}$ for a free wind. The numerical simulation of jets produced in type 
II collapsars, shown in fig. 11 of MacFadyen \etal (2001), shows that, after having propagated 
through the helium core, the angular distribution of the jet kinetic energy can be approximated 
as $dE/d\Omega \propto \theta^{-4}$ for $\theta \in (5^o,15^o)$. Such a steep fall-off of the 
ejecta kinetic energy away from the outflow axis corresponds to the J/O and J models.

 The third conclusion is that a spreading jet interacting with a homogeneous medium is the only 
viable model for the afterglows 990510, 000301, 011211, and 030226, as their break magnitudes $\Da$ 
are too large to be accommodated by a structured outflow (assuming the observer to be within the
opening of the core) or by a jet interacting with a free wind. In fact, for the afterglows 000301 and 
030226, the measured $\Da$ is so large, that even the J model is only marginally consistent with 
the observations. Larger break magnitudes would result from combining the EI and J models, though 
this would require a coincidence: the simultaneity of the energy injection cessation with the jet 
edge becoming visible to the observer. Stronger light-curve breaks would also result from structured 
outflows if the direction toward the observer were outside the core opening (Panaitescu \& Kumar 
2003), a case not included in this work. Such locations could also yield a cusp in the light-curve 
when the more energetic core becomes visible, \ie at the time of the light-curve break (Granot \& 
Kumar 2003).

 A more accurate comparison between the features of the above four models for light-curve breaks 
and the afterglow temporal/spectral properties requires numerical calculations of the outflow
dynamics and emission. Such calculations can follow the transition between the asymptotic power-law
light-curve decays and may rule out some of the models listed in Table 2, if this transition is 
too slow compared with the observations. Furthermore, numerical calculations provide the framework 
for inclusion of inverse Compton scatterings, which may affect the time-dependence of the cooling 
frequency (and, hence, the light-curve decay index at frequencies above it) and may contribute to 
the $X$-ray afterglow emission. These two factors have been left out in the derivation of the 
results presented in \S\ref{models}, \S\ref{criteria}, and \S\ref{tests}. Finally, numerical
modelling of the multiwavelength afterglow emission can determine other afterglow parameters
(ejecta energy, circumburst density, shock microphysical parameters) in addition to testing
the various possible mechanism for light-curve breaks. In this work we have described a simpler
way to test those mechanisms and have shown how some outflow properties (\eg energy distribution
with Lorentz factor or with polar angle) are constrained by afterglow observations.

\section*{Acknowledgments}
 This work was supported by NSF grant AST-0406878. The author is indebted to Pawan Kumar
 for helpful discussions and suggestions about this work.

\end{document}